\documentclass[conference]{IEEEtran}
\IEEEoverridecommandlockouts
\usepackage{graphicx} % Required for inserting images

\usepackage[style=ieee, backend=biber]{biblatex}
\usepackage{amsmath,amssymb,amsfonts}
\usepackage{algorithmic}
\usepackage{graphicx}
\usepackage{hyperref}
\usepackage{textcomp}
\usepackage{xcolor}
\usepackage{color}
\def\BibTeX{{\rm B\kern-.05em{\sc i\kern-.025em b}\kern-.08em
    T\kern-.1667em\lower.7ex\hbox{E}\kern-.125emX}}

\usepackage{flushend}
\usepackage{import}
\usepackage{tikz}
\graphicspath{{figures/}} %Setting the graphicspath

\ifCLASSOPTIONcompsoc
    \usepackage[caption=false, font=normalsize, labelfont=sf, textfont=sf]{subfig}
\else
\usepackage[caption=false, font=footnotesize]{subfig}
\fi
\usepackage[super]{nth}
\usepackage[inline]{enumitem}

\makeatletter
\newcommand{\linebreakand}{%
  \end{@IEEEauthorhalign}
  \hfill\mbox{}\par
  \mbox{}\hfill\begin{@IEEEauthorhalign}
}
\makeatother

\begin{document}

\title{Physical Attacks on the Railway System}
\author{\IEEEauthorblockN{Lukas Iffländer, Thomas Buder}
\IEEEauthorblockA{\textit{Research area Safety} \\
\textit{German Centre for Rail Traffic Research}\\
Dresden, Germany \\
\{IfflaenderL,BuderT\}@dzsf.bund.de} 
\and
\IEEEauthorblockN{Teresa Loreth, Marina Alonso Villota, Stefan Pickl}
\IEEEauthorblockA{\textit{Faculty of Computer Science} \\
\textit{Universität der Bundeswehr}\\
Munich, Germany \\
\{teresa.loreth,marina.alonso-villota,stefan.pickl\}@unibw.de} 
\and

\linebreakand

\IEEEauthorblockN{Walter Schmitz, Karl Adolf Neubecker}
\IEEEauthorblockA{%\textit{dept. name of organization (of Aff.)} \\
\textit{CreaLab GmbH}\\
Feldkirchen-Westerham, Germany \\
\{W.Schmitz,A.Neubecker\}@crealab-gmbh.de}
}
\maketitle

\begin{abstract}
    Recent attacks encouraged public interest in physical security for railways.
    Knowing about and learning from previous attacks is necessary to secure against them.
    This paper presents a structured data set of physical attacks against railways. 
    We analyze the data regarding the used means, the railway system's target component, the attacker type, and the geographical distribution of attacks.
    The results indicate a growing heterogeneity of observed attacks in the recent decade compared to the previous decades and centuries, making protecting railways more complex.
\end{abstract}

\section{Introduction} 
In October 2022, unknown attackers carried out an act of sabotage against the wireless communication infrastructure of Germany's primary railway infrastructure operator DB Netz AG~\cite{NoAuthor_2022_SabotageBahn}.
By selectively cutting cables at two distinct positions across Germany, the attackers rendered large parts of GSM-R (Global System for Mobile Communication---Rail), used for communication between train dispatchers and the trains themselves, unusable. 
In the following, traffic in northern Germany had to stop entirely until the system's repair.
This attack brought the vulnerability of the railway system into sharper focus.

The Russian invasion of Ukraine in the spring of 2022 also brings the vulnerability of railways further into the public eye---railway facilities are not only attractive military targets but also often targets for partisans.

Since the beginning of 2021, the project "Revealing Existing Attack Vulnerabilities in the Rail System" (REAVRS) has been running at the German Centre for Rail Traffic Research (DZSF) at the Federal Railway Authority. 
The contractors are the University of the Bundeswehr Munich and the Ingenieurgesellschaft für Verkehrs- und Eisenbahnwesen mbH (IVE mbH). 
The project deals with attack potentials on the railway system both on a physical and cyber level. 
One of the project's first goals was a historical analysis of physical attacks. 
This resulted in a systematically structured data set described and evaluated in this article\footnote{Data set will become publically available upon acceptance, reviewers find a temporary link at the end of the paper.}.
We considered the different types of attackers, the type of means of attack, the targets of the attacks and the damage incurred.
The results show, among other things, that the number of attacks has increased significantly over time and that there has been a diversification of attacks in the last decade. 
The security situation assessment is becoming more complex due to this increased diversity of attack characteristics.

The remainder of this paper is structured as follows: 
After this introduction, Section~\ref{sec:background}  briefly introduces relevant physical security aspects of the railway system.
Next, Section~\ref{sec:data set} describes the structure of our data set.
Building upon that, we analyze the data and present example attacks in Section~\ref{sec:results}.
We discuss these findings and their limitations in Section~\ref{sec:discussion}.
Furthermore, we take a look at related work in Section~\ref{sec:related}.
Finally, in Section~\ref{sec:conclusion}, we present an outlook on the further research work of the DZSF in this field and indicate the first trends regarding physical attack characteristics from the beginning of the current decade.

\section{Background}\label{sec:background}
The railway system is a critical infrastructure considered a powerful economic motor worldwide. It enables connectivity facilitating the transportation of goods and people while maintaining a lower environmental impact than other transportation systems. However, society's dependence on the railway system makes it an attractive target for hostile actors. The accessible designs of its infrastructure, both in the stations and in the railroads, make it vulnerable to physical attacks. In addition, the high numbers of people who use the railway daily, particularly the presence of central stations that gather thousands of people simultaneously, increase potential damages should an attack succeed. 

Despite the increasing reliance on public vigilance, CCTVs, and vigilant professionals at its stations, the railway system will never be completely secure against attacks. The composition of the railway system and overall design make it full of the so-called ‘soft targets’, places highly vulnerable but with a low level of protection~\cite{slivkova2023identification}. Thus, some criteria make specific nodes of the railway system more attractive than others to potential attacks: a high concentration of passengers in a given place, its centrality regarding the whole railway network (e.g., railway stations in capital cities and railway hubs), the elements on its vicinity (e.g., shopping and cultural centers, companies), and the presence of transport infrastructure of connecting operators in the vicinity of the station, such as terminals, other transportation stations, and stops~\cite{slivkova2023identification}. 

The destruction of railroads and railway infrastructure may give an advantage to an attacker in a warfare scenario, as it hinders the provision of resources and personnel, evacuations, and fast mobility \cite{Klein2022}; terrorist attacks on the railway system primarily aimed at disrupting services and/or causing casualties \cite{jenkins2020sophisticated}; and vandalism acts, such as the use of graffiti, cause a negative brand and reputation impact as well as considerable economic losses \cite{thompson2012broken}.

\section{Definition of Data set Structure}
\label{sec:data set}

In our data set, we created a structured representation of the attacks.
Each attack has the following properties, which themselves often comprise several sub-properties:

\begin{itemize}
    \item \textbf{Date:}
        When attacks lasted multiple days or were around midnight, we used the first of the relevant dates. 
        Attack dates resemble local times. 
    \item \textbf{Location:}
        The location specifies the country and the city in which the event occurred. 
        If available, we also specified the state.
        Furthermore, we added the latitude and longitude of the event location.
        In cases where the sources gave imprecise locations (e.g., only the line segment between two stations), we used applicable simplification (e.g., for the previous example, we used the middle between both stations).
        Lastly, when available, we also collected the name of the line of the attack. 
        When no line attribution was possible, we specify, e.g., the station or the attacked railway subsystem.
    \item \textbf{Attacker:}
        The attacker first comprises the type.
        We discern three types: \begin{enumerate*}
            \item state actors,
            \item non-state actors, and
            \item unknown actors
        \end{enumerate*}.
        When dealing with non-state actors, we consider whether states back their activities.
        For known actors, we further specify whether we deal with individual or group actors.
        Lastly, we note the name or alias of the actors.
    \item \textbf{Attack Target}:
        The attack target describes the intended target for the attack.
        We discern seven attack target types:
        \begin{itemize}
            \item \emph{Tracks:} 
                Tracks comprise the rails themselves, sleepers, fastenings, and sub- and superstructures. 
            \item \emph{Bridges:}
                We consider bridges a target if attacks target the bridge directly (e.g., explosives at support pillars). 
                Attacks that target the tracks on a bridge (e.g., by loosening fasteners) fall under the category of \emph{Tracks}.
            \item \emph{Stations:}
                In this category, we consider attacks at and in station buildings.
            \item \emph{Rolling Stock:}
                This category comprises attacks that target rolling stock (e.g., bombs in a carriage).
            \item \emph{Passengers:}
                For this category, we look at attacks that directly attack passengers (e.g., knife attacks on passengers).
            \item \emph{Communication:}
                These attacks comprise any attacks against communication infrastructure (e.g., taking out railway wireless communication) and signaling and interlocking components (e.g., setting fire to cables used for interlocking).
            \item \emph{Other:}
                This last category comprises all targets not covered by the other categories. 
                A real-world example would be an attack on a railway authority building. 
        \end{itemize}
        Multiple events lack the specification of the concrete attack target in the available sources. 
        Some events have multiple attack targets (e.g., bombs exploding in trains exactly when they pass through stations)
    \item \textbf{Means of Attack:}
        Similar to the attack targets, we tracked the means of attack. We discern the following means of attack:
        \begin{itemize}
            \item \emph{Explosives}
            \item \emph{Knife or Baton}
            \item \emph{Obstacles}
            \item \emph{Fire:} 
                This category excludes secondary fires, e.g., explosions.
            \item \emph{Rail Manipulation:}
                This category excludes secondary rail manipulations, e.g., broken rails from an explosion.
            \item \emph{Firearms}
            \item \emph{Suicide Bomber}
            \item \emph{Gas}
            \item \emph{Missiles}
            \item \emph{Hacker:}
                This category applies to physical attacks supported by hacker attacks and does not comprise pure hacker attacks
        \end{itemize}
        Again, multiple events either do not indicate the used means of attack, and others have more than one means of attack. 
    \item \textbf{Impact:}
        This category describes the actual impact.
        Note that attacks can target one part of the railway system but have effects on other parts or miss the effect entirely.
        For example, a sabotaged bridge can collapse under a train and cause casualties and damage to the rolling stock.
        Otherwise, loosening fasteners targets the tracks but could be detected and not cause any impact.
        We discern multiple impact types:
        \begin{itemize}
            \item \emph{Infrastructure:} 
                We summarize all significant damage to tracks, bridges, and stations under infrastructure. 
            \item \emph{Rolling Stock:}
                Whenever a rolling stock is significantly damaged, it falls in this category.
            \item \emph{Injured:}
                Here, we collect the number of injured people.
                If possible, we discern between light and severe injuries. If not, we consider all reported injuries as light. 
                We use the lowest number when the sources provide ranges of injured casualties. 
            \item \emph{Dead:}
                This category holds the number of deceased people.
                We use the lowest number when the sources provide ranges of deceased. 
        \end{itemize}
    \item \textbf{Description:}
        This last item contains the textual description of the event.
\end{itemize}

\begin{figure}[tb!]
    \centering
    \input{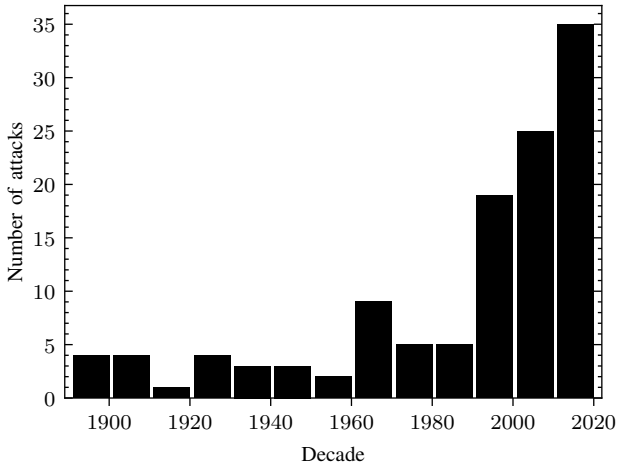}
    \caption{Development of the number of physical attacks per decade. Attacks up until 1900 are summarized as one bar.}
    \label{fig:attacks_by_year}
\end{figure}

\begin{figure}[tb!]
    \centering
    \input{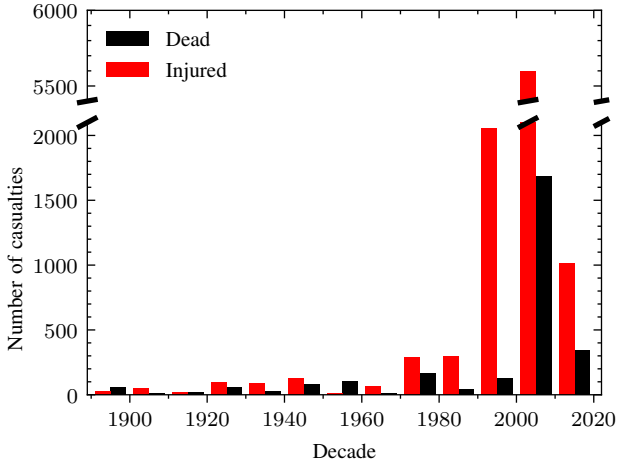}
    \caption{Development of the number of casualties per decade separated between injured and dead. Attacks up until 1900 are summarized as one bar.}
    \label{fig:cas_by_year}
\end{figure}

\begin{figure}[t!]
    \centering
    \input{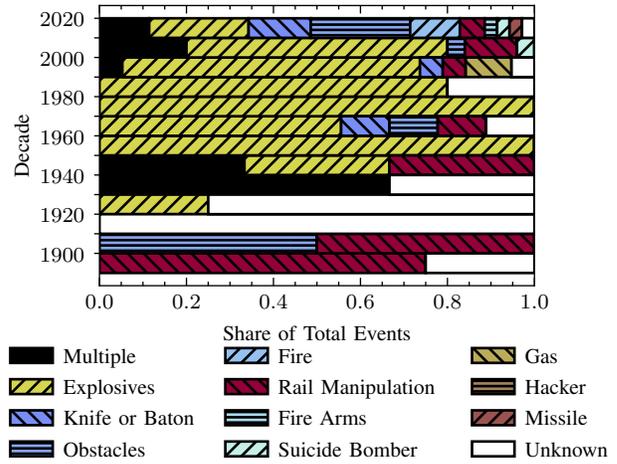}
    \caption{Share of different means of attacks of the total events in a decade. Events with more than one mean are described as `Multiple' while events without a known mean are described as `Unknown.'}
    \label{fig:means_share}
\end{figure}

\begin{figure*}[tb!]
    \centering
    \subfloat[Outliers included\label{fig:cas_by_event:outliers}]{%
        \input{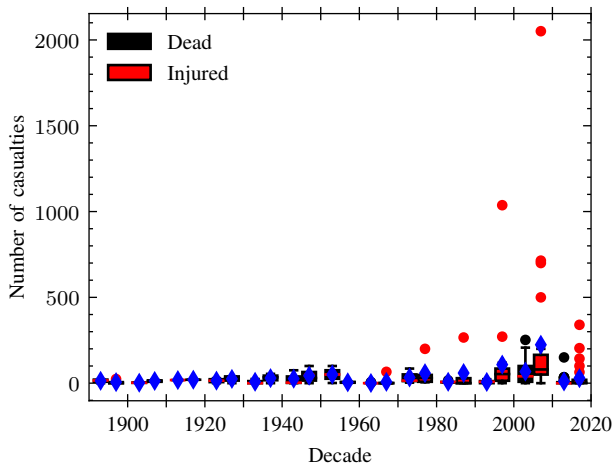}}
    \hfill
    \subfloat[Outliers excluded\label{fig:cas_by_event:pruned}]{%
        \input{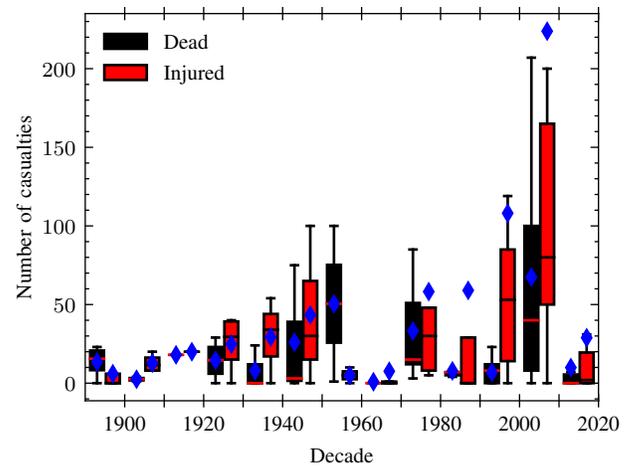}}
    
    \caption{Development of the number of casualties per event separated between injured and dead. Attacks up until 1900 are summarized as one bar. Since the whiskers for the outliers massively reduce the readability, we show a version of the same data with outliers on the left in Subfigure~\ref{fig:cas_by_event:outliers} and without on the right in Subfigure~\ref{fig:cas_by_event:pruned}}.
    \label{fig:cas_by_event}
\end{figure*}

\section{Results and Data aggregation}\label{sec:results}

In total, we collected 127 events.
We deduced trends and insights into the characteristics of the collected physical attacks using descriptive analysis. 
Even though we collected data past 2020, we neglected attacks from 2020 onward in our analysis.
We made this decision because many attacks are not yet solved, and, e.g., the identity of an attacker is not yet known but might be discovered shortly.
Furthermore, many events have not been collected. Especially in countries that significantly limit the freedom of the press, the reporting on such events can be largely delayed.
We will, however, take a short look at the events after 2020 in the outlook part of the conclusion in Section~\ref{sec:conclusion}.

\subsection{Overview of Results}

We first examined the development of the number of attacks carried out over time. As Figure \ref{fig:attacks_by_year} indicates, the number of attacks has risen significantly since the 1990s. This trend continued in the following decades, reaching a maximum of 35 documented attacks in the 2010s. 

A similar result holds for the analysis of the number of casualties; see Figure \ref{fig:cas_by_year}. The analysis of the casualties per event depicted in Figures \ref{fig:cas_by_event:outliers} and \ref{fig:cas_by_event:pruned} yields that also the average impact of the attacks increased since the last decades of the \nth{20} century. This is particularly evident because the upward outliers regarding the number of injured persons mainly appear only since the 1980s (Figure \ref{fig:cas_by_event:outliers}).
The high results from the 2000s are driven by multiple high-casualty events, e.g., the Madrid bombings in 2004 (1\,969 lightly and 82 severely injured and 192 dead) or the 2006 Mumbai bombings (714 injured and 207 dead).
The difference between the figures' median and average visualizes this effect.

Next, Figure~\ref{fig:means_share} shows the development of the share each mean of attack has through decades.
In the beginning, most attacks used either obstacles or the manipulation of rails.
Especially in the early decades, we found multiple imprecise sources that did not give the used mean of attack. 
From the post-war (after World War~II) period on, most attacks employed used explosives.
This trend held until the second decade of the \nth{21} century when attacks with melee weapons, obstacles, and fire took a significant share. Moreover, the distribution of means is generally more heterogeneous.
It is worth mentioning that in this last decade, the first missile attack happened. 
Also, suicide bombers started to appear at the turn of the century. 

Figure~\ref{fig:attackers_share} shows the composition of the attackers.
We find that in the early decades, most attackers are unknown. 
This changed around the mid of the \nth{20}-century. 
This fact is due to better documentation of state-backed attacks and the strife from terrorists for recognition of their feats.
Except for the 1940s and 1950s, most known attacks come from non-state actors. 
Also, in most decades, the majority of attacks come from groups. 
Notable exceptions are the 1960s due to multiple attacks in Germany from Alexander Bordan Hembluck and the 2010s from a single Daesh supporter, also in Germany.
State-backed non-state actors are rare and only comprise Confederate guerrillas in the American Civil War, the Irish Republican Army, and a group backed by the Pakistani secret service.

\begin{figure}[t!]
    \centering
    \input{figures/share_of_attackers_by_year.pgf}
    \caption{Share of different attacker types of the total events in a decade. Events without a known attacker are described as `Unknown.' The coloring represents state or non-state actors while the hatches indicate groups and state-backing.}
    \label{fig:attackers_share}
\end{figure}

\begin{figure}[tbh!]
    \centering
    \input{figures/share_of_attack_targets_by_year.pgf}
    \caption{Share of different attack targets of the total events in a decade. Events with more than one target are described as `Multiple' while events without a known mean are described as `Unknown.'}
    \label{fig:targets_share}
\end{figure}

As a last distribution, we consider the various attack targets.
Figure~\ref{fig:targets_share} visualizes these results. 
In the first centuries, most attacks with known targets mainly focused on the infrastructure side (tracks and bridges). 
The outlier in the 1910s results from this decade only comprising a single attack.
From the 1960s, the share of attacks on rolling stock grew and became the majority of known attack targets until the 2010s.
Also, post-war, attacks against stations and communication infrastructure emerge.
As for the means of attack, the 2010s show a rather heterogeneous composition.

\begin{figure*}[bth!]
    \centering
    \input{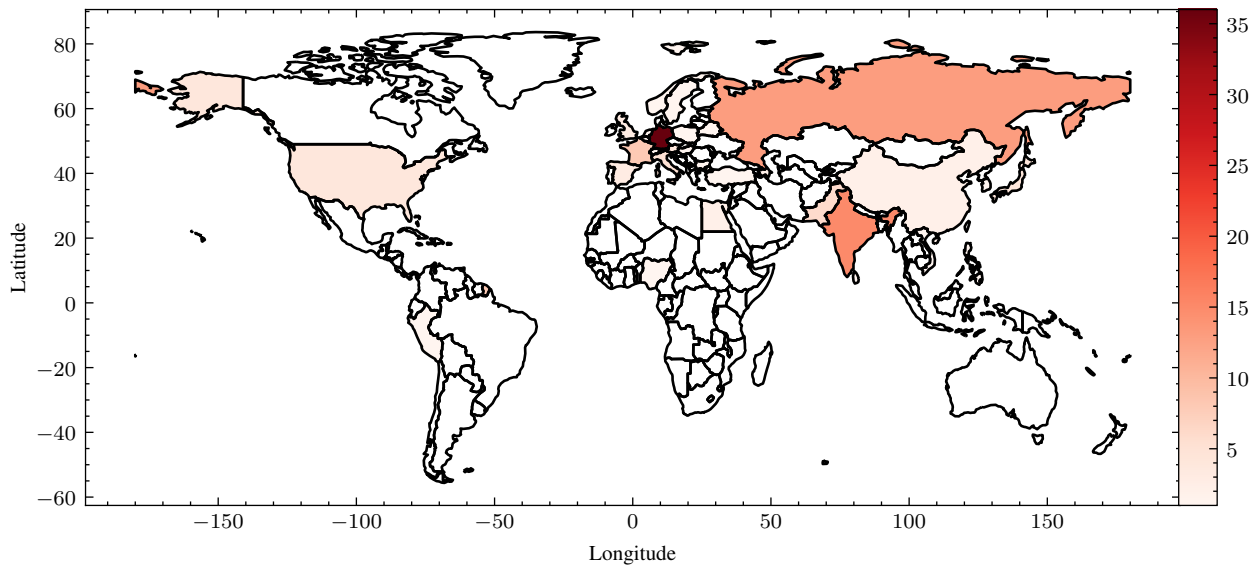}
    \caption{The number of attacks per country since the invention of railways.}
    \label{fig:attacks_by_country}
\end{figure*}

Besides the distribution over all attacks regarding the target, means, and attacker type, we also considered the geographical distribution. 
Figure~\ref{fig:attacks_by_country} gives a world map comprising all countries and the number of attacks.

The map shows that most attacks occurred in Germany.
Further points of frequent attacks are Russia and India.
In general, the majority of attacks happen in Europe and Asia.

\begin{figure*}[htb!]
    \centering
    \input{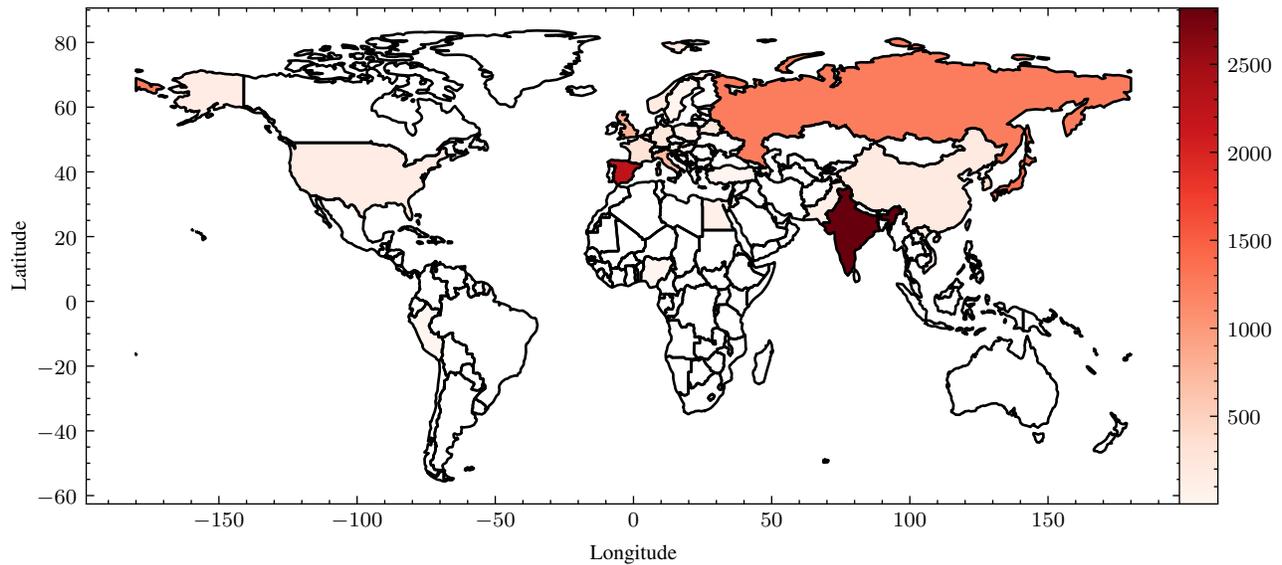}
    \caption{The amount of casualties per country since the invention of railways. The given amount is the sum of deceased and injured persons.}
    \label{fig:casualties_by_country}
\end{figure*}

Further, Figure~\ref{fig:casualties_by_country} presents the casualties per country on a world map. 
While the number of attacks might indicate that Germany would also suffer from many casualties, the opposite is the case.
We attribute the difference between attack frequency and impact to many sabotage events, especially from left-wing extremists.
While Russia and India have many attacks and casualties, Spain and Japan have high casualties with fewer attacks.
This effect results from very few high-intensity attacks. 
Again the majority of casualties come from Europe and Asia.

\subsection{Representative Attack Examples}

In this section, we want to give some representative examples of observed attacks.
Due to space limitations, we can not give a complete rundown of every event.

\subsubsection{\nth{29} of October 1888, Russia, Birky}

An example of an event with little available data is the attack on the \nth{29} of October 1888. 
On this date, the train of Tsar Russian Alexander III derailed. 
Russian authorities considered it to be an attack.
Casualties comprised 24 injured and 23 dead.
No further information on this attack is available.
Thus, we considered the attacker, the attack target, and the used means as unknown~\cite{Stockert_1913_EisenbahnunfalleEinBeitragZurEisenbahnbetriebslehre}.

\subsubsection{\nth{9} of September 1991, India, Mumbai}

This example is typical for many attacks in the 1990s, resulting in a mid-two-digit number of injured (60) and a small two-digit number of dead (10) people.
On the \nth{9} of September 1991, an attack was carried out by Islamist terrorist groups on a suburban train in Mumbai, India.
Here, the sources~\cite{Chaba_2016_Kalyan1991TrainBlastSITtoProbeConvict’sActivitiessinceHeJumpedParolein2007,Sabharwal_2002_RavinderSingh@BittuVstheStateofMaharashtraon30April2002,Sehgal_2016_1991MumbaiBlastConvictArrestedfromPhagwara} allowed to designate the rolling stock as the target and explosives as the mean used.

\subsubsection{\nth{11} of March 2004, Spain, Madrid}

The attack in Madrid is relevant as a maximum casualty event, with 191 dead and 2021 injured. 
Here the sources allow for a detailed accounting of the relevant information with a minute-by-minute description.
On the \nth{11} of March 2004, between 7:39 and 7:42, ten bombs simultaneously exploded in four fully occupied suburban trains and at Atocha station.
Thus, we marked rolling stock and the station as the attack targets and explosives as the primary mean of attack. 
Furthermore, reports accredit responsibility to a terror cell financed by al-Qaida. 
Therefore, we marked the attacker as a non-state actor group without state support~\cite{Universal_2006_MadridBombingProbeFindsNoAlQaidaLink}.

\subsubsection{\nth{29} of March 2010, Russia, Moscow}

An attack close to the median casualties of its decade occurred on the \nth{29} of March 2010 in Moscow.
That day, 100 people were injured (12 light and 88 severely), and 40 died.
Two explosions occurred on the platforms of metro stations, thus, targeting the stations and the trains on the platform using the means of explosives.
The attack was credited to the Chechen Black Widows, a non-state group actor without state support~\cite{NoAuthor_2012_KaukasischeTerrorgruppenHinterAnschlaegenAufUBahnVermutet,NoAuthor_2023_BBCNewsMoscowMetroHitbyDeadlySuicideBombings}.

\subsubsection{\nth{19} of June 2017, Germany, nationwide}

A last example is an attack without casualties.
On the \nth{19} of June 2017, 13 arson attacks were carried out simultaneously on cable installations of railway infrastructure in five German states, causing significant damage and severely degrading signaling and communication.
Thus, the mean was fire, and the target was signaling and communication infrastructure.
Left-wing extremists claimed responsibility for the attacks, leading to a classification as a non-state actor without state support.
Furthermore, due to the simultaneity of the attack, we consider the actor to be a group actor~\cite{NoAuthor_2017_AnschlaegeAufKabelanlagenDerDB}.

\section{Discussion and Limitations}\label{sec:discussion}

Our research suggests that the total number of attacks and the average impact per attack have increased, especially since the end of the last century. The means of attack and targets get more heterogeneous. This increased diversity of attack properties requires a holistic, broad strategy for suitable safety measures. Moreover, the recent attack on the communication infrastructure in Germany indicates a better system knowledge of potential attackers. However, more classical means of attack, such as knives or fire, still play a significant role in recent days.
\par
The derived quantitative estimates of attack trends and characteristics naturally depend on the search strategy and defined data structure. This results in two main reasons for the incompleteness of the researched attacks. First, we only searched in English and German, so we missed attacks only reported in other languages. However, the distribution of the researched attacks covers the whole world, indicating that we could cover at least a collection of the most relevant attacks. Second, we focused on digital references, which may have biased the number of attacks covered in earlier years. However, where available, we incorporated older text sources like~\cite{Stockert_1913_EisenbahnunfalleEinBeitragZurEisenbahnbetriebslehre,NoAuthor_2017_AnschlaegeAufKabelanlagenDerDB}. This suggests that our analysis underestimates the number of attacks until the late 1990s. Despite this limitation, the references suggested we covered the most significant attacks in this period. Furthermore, the general inferred trend of an increase in the number of attacks during the 2000s holds.

\section{Related Work}\label{sec:related}

Concerning the related work of our quantitative analysis, a few other publications should be mentioned in this context. Hartong et al.~\cite{Hartong_2008_SecurityandtheUSRailInfrastructure}, Sarkar \& Sarkar~\cite{Sarkar_2022_TrainDisastersintheHistoryofIndianRailwaysaReview} and De Cillis et al.~\cite{Cillis_2013_AnalysisofCriminalandTerroristRelatedEpisodesinRailwayInfrastructureScenarios} were therefore identified. They all provide an overview of developments of physical attacks and incidents in railway infrastructures. 

In Hartong, et al.~\cite{Hartong_2008_SecurityandtheUSRailInfrastructure} deal with the significant role of railroads in the United States and highlights its weaknesses regarding attacks and incidents that occurred. By analyzing attacks characterized by a more straightforward structure that led to the disruption of the entire railway system, the importance of the passenger rail in the United States is the main focus throughout the paper. In conclusion, however, even if holistic protection against all kinds of attacks is not a realistic scenario, concrete measures to take for reducing the risk of attacks by analyzing the vulnerabilities of the entire railroad security system are represented. 

Sarkar \& Sarkar present a study~\cite{Sarkar_2022_TrainDisastersintheHistoryofIndianRailwaysaReview} which is akin to our analysis. It shows how the number of injured and deaths in Indian railway accidents (natural- and man-made) varies from the mid-20th century to 2022. In general, the number of injured and dead people has declined in recent decades. In particular, during the 70ies, railway accidents were at their lowest point since 1947. However, from the 90ies to 2020, the number of accidents that were provoked by natural- (17\%) and man-made causes (17\%), e.g., terrorism, as well as by technology and mechanics (66\%), is steady. While the number of accidents has been constant for the last decades, the number of injured and killed people is relatively high between 1981 and 2010. A reason for this is seen in the growing Indian population. From 2011 until 2022 the number of humans negatively affected declined because of technological- and medical improvements. 

Cillis et al. provide a third publication related to our work~\cite{Cillis_2013_AnalysisofCriminalandTerroristRelatedEpisodesinRailwayInfrastructureScenarios}. The data collected by identifying around 540 attacks initiated by terrorist and security incidents in international railways between the early 70's to 2011 is saved in a specifically created open-source-based database called RISTAD (Railway Infrastructure Systems Terrorist Attacks). By applying the database, an analysis of the counterintuitive correlations between the attacks and the characteristics of an object becomes visible, for instance, the illustration of the lethality on the one side while showing the number of tracks available within a particular station on the other side. The general focus of this research lies in identifying railroad facilities related to infrastructural or environmental facets, which make a target more attractive to attack.

\section{Conclusions and future work}
\label{sec:conclusion}

In this paper, we collected data about physical attacks on railway infrastructure in a structured data set. 
We analyzed the collected data concerning means of attack, attacker types, attack targets, casualties and global distribution.
Our findings allow us to discern trends. Especially the continuous growth in total attacks over multiple decades leads to the expectation that even more attacks might occur in the future.  
Also, especially in the 2010s, the means and targets of the observed attacks became more diverse.
Thus, it is no longer possible to focus solely on protecting against attacks with explosives like in the previous decades, but instead, it becomes necessary to anticipate more complex attacks.
The casualty data gives an optimistic lookout, since the number of casualties vastly shrank in the 2010s compared to the previous two decades. 
Combined with the knowledge about more frequent attacks, this leads to the assumption, that attacks are either becoming less lethal or security mechanisms become efficient enough to better prevent attacks like observed in the 1990s and 2000s.

The recording of attacks is part of the first work package of the REAVRS project. Further packages deal with the definition of relevant attack characteristics, their threat assessment, and the derivation and evaluation of countermeasures. In addition, the project also includes the recording and analysis of cyber attacks, which will be carried out in a methodologically similar way. 

The DZSF plans to make the described data set available on its website. Furthermore, in the project context, the possibility is to be created to report errors in the data set, suggest corrections, and contribute new attacks. It is being examined how this can be mapped within the current and future IT infrastructure of the Federal Railway Authority. 

Last, we take a look at the attacks of the current decade. In that case, it is noticeable that the attacks increasingly relate to communication and facilities of the control and safety technology and weaken the system's availability. 
Since technology development is advancing rapidly in this area in particular, the DZSF is currently investigating which IT security concepts will be necessary for the future and suitable for adequately protecting rail transport against cyber attacks in another project entitled "Forecast of security requirements and evaluation of possible security concepts for the railway system." The report on the technology forecast has already been published \cite{DZSF_Prognose}.

\section*{Data Availability}

We provide the data set in the form of an Excel sheet under~\cite{ifflander_physical_2023}.
Future updates of the data set will be published on the same platform.

\section*{Acknowledgement}

We want to thank our DZSF student interns Henriette Kobl and Mark Essegern for their support in migrating the data from plain text to structured lists.

%\begin{figure*}[tb!]
%    \centering
%    \subfloat[Outliers included\label{fig:cas_by_cas_event:outliers}]{%
%        \input{figures/average_casualties_by_year_at_casualty_events.pgf}}
%    \hfill
%    \subfloat[Outliers excluded\label{fig:cas_by_cas_event:pruned}]{%
%        \input{figures/average_casualties_by_year_at_casualty_events_no_whiskers.pgf}}
%    
%    \caption{Development of the number of casualties per event that had casualties of that type separated between injured and dead. Attacks up until 1900 are summarized as one bar. Since the whiskers for the outliers massively reduce the readability, we show a version of the same data with outliers on the left in Subfigure~\ref{fig:cas_by_cas_event:outliers} and without on the right in Subfigure~\ref{fig:cas_by_cas_event:pruned}}.
%    \label{fig:cas_by_cas_event}
%\end{figure*}

%\section*{Bibliography}

\printbibliography

\end{document}